# Relativistic Shear Flow Between Electron-Ion and Electron-Positron Plasmas and Astrophysical Applications

Edison Liang[1], Wen Fu[1], Markus Böttcher[2]

[1] Rice University, Houston, TX 77005

[2] North-West University, Potchefstroom, 2520, South Africa

## Abstract

We present Particle-in-Cell simulation results of relativistic shear boundary layers between electron-ion and electron-positron plasmas and discuss their potential applications to astrophysics. Specifically, we find in the case of a fast electron-positron spine surrounded by a slow-moving or stationary electron-ion sheath, lepton acceleration proceeds in a highly anisotropic manner due to electromagnetic fields created at the shear interface. While the highest-energy leptons still produce a beaming pattern (as seen in the quasi-stationary frame of the sheath) of order $1/\Gamma$, where $\Gamma$ is the bulk Lorentz factor of the spine, for lower-energy particles, the beaming is much less pronounced. This is in stark contrast to the case of pure electron-ion shear layers, in which anisotropic particle acceleration leads to significantly narrower beaming patterns than $1/\Gamma$ for the highest-energy particles. In either case, shear-layer acceleration is expected to produce strongly angle-dependent lepton (and hence, emanating radiation) spectra, with a significantly harder spectrum in the forward direction than viewed from larger off-axis angles, much beyond the regular Doppler boosting effect from a co-moving isotropic lepton distribution. This may solve the problem of the need for high (and apparently arbitrarily chosen) minimum Lorentz factors of radiating electrons, often plaguing current blazar and GRB jet modeling efforts.

Subject Keywords: Shear Flow; Gamma-Ray Bursts; Quasars

# 1. INTRODUCTION

Recent large-scale Particle-in-Cell (PIC, Birdsall & Langdon 1991) simulations of relativistic collisionless shear boundary layers (SBLs) demonstrated efficient generation of large-scale transverse magnetic fields and particle energization from initially unmagnetized plasmas (Alves et al 2012, 2014, Grismayer et al 2013, Liang et al 2013ab). Relativistic SBLs are likely to arise from the interaction of relativistic outflows with slow-moving or stationary ambient medium, from blazar jets (Böttcher 2007, Ghisellini et al 2005), gamma-ray bursts (GRB, Piran 2004, Mészáros 2002) to pulsar winds (e.g. Kargaltsev et al. 2015). Recent data modeling suggests that some blazar jets may have a spine-sheath structure (Ghisellini et al 2005, 2010) which also solve certain problems of single-component jets (Maliani and Keppens 2009, Lyutikov and Lister 2010), and enhanced emission at the spine-sheath boundary may be the source of the observed limb brightening (Girolette et al 2004). Previous PIC simulations of SBLs focused on shear flows in which the plasma composition is the same on both sides of the shear boundary (e.g. pure electron-positron (e+e-) plasma, Liang et al 2013a, pure electron-ion (e-ion) plasma, Alves et al 2012, Grismayer et al 2013, or uniform mixtures of e+e- and e-ion plasma, Liang et al 2013b). In this paper we extend the previous works to include jumps in the plasma composition across the shear boundary. Specifically we consider the case where the plasma is pure e+e- on one side of the SBL and pure e-ion on the other side of the SBL. This case is most relevant to

GRB jets and pulsar winds where the relativistic outflow from the central engine is conjectured to be e+e- plasma, while the ambient CSM or ISM is purely e-ion plasma, but it may also be applicable to some blazars. We will first summarize the PIC simulation results in Sec.2, and then discuss potential astrophysical applications in Sec.3. As in our previous PIC simulations (Liang et al 2013ab), we work in the frame in which the two opposing shear flows have equal Lorentz factors (Fig.1), which was previously called the center of momentum (CM) frame in the cases of homogenous composition. However, in the present case, since the two opposing flows have very different masses, this frame of *equal-Lorentz-factors (ELF) is not the CM frame*. All results presented below can be Lorentz boosted to the "laboratory" frame where the ambient e-ion sheath is initially at rest, or to the "comoving" frame of the outflow, where the e+e- spine is initially at rest.

First we briefly summarize previous PIC simulations results on SBLs with homogenous composition. In the pure e+e- case (Liang et al 2013a), transverse electromagnetic (EM) fields are first created by the oblique Weibel (Weibel 1959, Yoon 2007, Yang et al 1993, 1994) and 2-stream (Boyd & Sanderson 2003, Lapenta et al 2007) instabilities caused by interface-crossing particle streams. These small-scale fields then grow into larger organized quasi-periodic magnetic flux ropes with alternating polarity. At late stage, they coalesce into large nonlinear eddy-like EM "vortices" extending up to hundreds of electron skin depths (Fig.2f, Liang et al 2013a). Particles are energized by **ExB** drift motions (Boyd & Sanderson 2003). Eventually a steep power-law tail is formed which turns over at $\sim p_o/2$, where $p_o$ is the initial bulk Lorentz factor in the ELF frame. A key feature is the formation of a density trough at the SBL due to plasma expulsion by the self-created magnetic pressure. The EM energy saturates at $\sim 8\%$ of the total energy, which is dominated by the longitudinal P-mode at early times, but is dominated by the transverse T-mode at late times (Fig.1, Liang et al 2013a).

In the pure e-ion ($m_i/m_e$=1836) case (Alves et al 2012, 2014, 2015, Grismayer et al 2013), the development and structure of the SBL is completely different from the e+e- case. Due to their large mass difference, the ions are free-streaming while the electrons are fluid-like (Gruzinov 2008, Grismayer et al 2013). For cold electrons the initial instability is a kinetic version of the Kelvin-Helmohltz instability (KHI, Chandrasekhar 1981, Zhang et al 2009) involving only the electron fluid. Once the electrons achieve finite transverse momentum, either due to finite temperature or the Gruzinov instability, the electron counter-current instability (ECCI, Fig.1 inset, Liang et al 2013b, Alves et al 2012) dominates as electrons with finite transverse momentum cross over to the opposite flows, while the heavy ions do not. This leads to the formation of opposite current layers on the two sides the shear interface, creating a monopolar slab of dc magnetic field (Fig.2d, Alves et al 2012, Liang et al 2013b). The dc magnetic pressure expels the plasma, forming an effective ion vacuum gap at the interface (Liang et al 2013b). But the electrons, which have lower mass, are less evacuated than the ions. This leads to charge separation and formation of a triple layer (i.e. double capacitor), with strong $\mathbf{E}_y$ fields pointing from both current layers towards the center of the SBL (Liang et al 2013b). At the same time, $\mathbf{E}_x$ fields (parallel to the current in the direction of the bulk motion) induced by $\partial \mathbf{B}_z/\partial t$ are generated outside the current layers, which efficiently accelerate electrons towards the ion energy. Eventually the electron distribution forms a narrow peak at $\sim p_o m_i c^2/2$. Unlike the pure e+e- case, there is no evidence of power-law tail formation. The EM energy saturates at $\sim$ 12% of the total energy and is completely dominated by the longitudinal P-mode. The transverse T-mode saturates at very low level, and contributes little to field creation or particle energization.

When we add e+e- plasmas uniformly to the e-ion shear flow, the results become more interesting. Besides the slab dc fields and ion vacuum gaps created at the interface by the ECCI, large-amplitude EM waves are created just outside the current layers, sandwiching the slab dc fields (Fig.2e, Liang et al 2013b). In addition to coherent acceleration by the dc fields towards the ion energy, leptons are also stochastically scattered by these EM waves to form a power-law tail, which at late times approaches a slope of ~ -3 (Liang et al 2013b). Hence homogeneous e+e-ion shear flows may potentially explain the observed gamma-ray spectra of GRBs and blazars. Figure 2 summarizes the previous results of the three homogeneous shear flows discussed above (see Liang et al 2013ab for details). A comment is in order here on the effect of the initial thickness of the velocity transition layer. While the steepness of the initial velocity gradient affects the linear growth rate of the ECCI, it does not affect the nonlinear saturation of the EM energy or the final structure of the SBL, as long as the initial velocity transition layer is thinner than the final thickness of the SBL (Fig.2).

## 2. PIC SIMULATION RESULTS

In the current runs, we used both the 2.5D (2D space, 3-momenta) code Zohar (Birdsall & Langdon 1991, Langdon & Lasinski 1976) and the fully 3D code EPOCH (Brady et al 2012). As in other ion-dominated shear flows, the T-mode in the y-z plane (Fig.1) saturates at low amplitude and has negligible effect in this case (Liang et al 2013b, confirmed by both 2.5D runs in the y-z plane and 3D runs). Hence we focus only on the P-mode (Fig.1) and present only the 2D results in the x-y plane, since we can use much larger simulation boxes in 2D and run for longer times without any (artificial) interaction between the two SBLs (Fig.1). To compare with the previous results (Fig.2), in this section we present the results in the ELF frame for initial shear Lorentz factor $p_o$=15 and initial temperature $kT_i$ = 2.5 keV for electrons, positrons and ions ($m_+/m_e$=1, $m_i/m_e$=1836).  In the ELF frame the ions carry almost all of the initial energy. Throughout this paper and in all figures, distances are measured in $c/\omega_e$ ($\omega_e$ = electron plasma frequency) and times are measured in $1/\omega_e$.  We normalize the initial density n=1 for all three species so that the cell size = electron skin depth.  The plasmas are initially unmagnetized. Initially right-moving e+e- plasma occupies the central 50% of the y-grid (hereafter called the "spine"), while left-moving e-ion plasma occupies the top 25% and bottom 25% of the y-grid (hereafter called the "sheath").  We call this configuration the e+e-ion sandwich (Fig.1).  To study the effect of grid size, we varied the y-grid from 2048 cells up to 12288 cells. The number of super-particles per cell varied from 10 to 20 for each species.  To maximize numerical stability, we used time-step $\Delta t = 0.1/\omega_e$ in all runs.  For the 1028 x 12288 EPOCH run, we ran to > 400000 time steps.  Overall energy conservation was better than 1% for the Zohar code and was ~ 5% for the EPOCH code.  We found little difference between the main results of the two codes for the same setup.

Below we highlight the major results of the e+e-ion sandwich SBL in the ELF frame. Figures 3-6 show results of 1024x2048 runs, while Fig.7 shows results of a 1024x12288 EPOCH run. Figure 3 shows the energy flow between ions, leptons and EM fields. We see that the electron energy saturates at ~25% of total energy, positron energy saturates at ~ 4%, and EM energy saturate at ~ 3% of total energy, much lower than the homogeneous cases of Fig.2.  (The inset shows that EM energy growth in the T-mode is negligible.)  Figure 4 shows the $\mathbf{B_z}$ $\mathbf{E_y}$, $\mathbf{J_x}$, $\mathbf{E_x}$ and net charge $\rho = (n_+-n_-)$ profiles at $t\omega_e$=3000, 10000, 16000  respectively.  Comparing with Fig.2, we see that the e+e-ion sandwich SBL is a hybrid of the pure e-ion SBL and the pure e+e- SBL:

the e-ion half of the SBL is dominated by strong persistent monopolar fields, while the e+e- half of the SBL contains bipolar EM waves. Global inductive $\mathbf{E_x}$ fields permeate both the e-ion and e+e- regions away at early times (Fig.4(a)). In the e-ion sheath, the $\mathbf{E_x}$ field points to the right, decelerating the protons while accelerating the electrons in the –$\mathbf{x}$ direction. In the e+e- spine, the $\mathbf{E_x}$ field points to the left, accelerating electrons in the +$\mathbf{x}$ direction and positrons in the –$\mathbf{x}$ direction. But at late times (Fig.4(b)(c)), only weak wave fields remain. Figure 5 shows the evolution of the x-averaged densities of ion, electron, positron and net charge as functions of y. This shows that an electron-dominated negative layer persists on the e+e- side of the SBL, while an ion-dominated positive layer persists on the e-ion side of the SBL, creating a double layer instead of the triple layer of the e-ion case (Fig.2). The currents associated with the charge separation sustain the dc slab field at the interface (Fig.4). Figure 6 shows the time evolution of the electron, ion and positron energy spectra. At late times the total electron spectrum consists of a quasi-power-law extending from $\gamma_{min} \sim 400$ to $\gamma_{max} \sim 2 \times 10^4$ (=peak ion energy), plus a narrow peak around $\gamma \sim 10^4$. Detailed analysis suggests that the power-law electrons come from the e+e- spine, while the narrow peak population come from the e-ion sheath. The positrons spectrum is much softer than the electrons. Ions are continuously decelerated over time, peaking at $\gamma_i \sim 10$ at late times with a broad low-energy distribution (Fig.6c). In short, the e+e-ion sandwich SBL is capable of accelerating electrons to form a power-law tail extending to the ion peak energy ($\gamma_e \sim$ 20000 or 10 MeV, Fig.6a). Figure 7 shows the SBL evolution for our large 1024 x 12288 EPOCH run. For this larger y-grid, EM fields originating from the two SBLs do not interact significantly until $t\omega_e \sim 15000$. At late times, nonlinear oblique EM waves dominate the spine region and the e+e- spine expands against the e-ion sheath due to pressure imbalance. Scattering by these EM waves plays an important role in forming the power-law electron population.

# 3. APPLICATIONS TO GRBS AND BLAZARS

If GRB and blazar jets indeed have a spine-sheath structure, then our PIC results above may be applicable to the local emission properties of the spine-sheath interface. In order to visualize the key differences in particle and radiation properties between a pure e-ion jet and a jet with e+e- spine surrounded by e-ion sheath, it is better to Lorentz boost the particle momenta from the ELF frame of Sec.2 to the "laboratory" frame (LF) *where the sheath is initially at rest, while the spine moves with bulk Lorentz factor $\Gamma$=451*. Figures 8, 9 compare the phase plots for the two cases, after Lorentz boosting (in the **−x** direction) from the ELF frame to the LF. We see that for the pure e-ion SBL (Fig.8), spine electrons are accelerated up to 0.5 TeV (~ decelerated ion energy in LF). Fig.8a shows that accelerated spine electrons reach extreme momentum anisotropy ($\mathbf{p_{xL}} \gg \mathbf{p_y}$), while their beam angle $|\mathbf{p_y/p_{xL}}|$ decreases with increasing energy (Fig.10a). On the other hand, for the e+e-ion sandwich SBL (Fig.9), the spine leptons are accelerated only up to ~ 10 GeV (~ accelerated ion energy in LF). While the accelerated spine lepton momenta are still anisotropic (Fig.9a), their beam angles are now much broader (Fig.10b). In fact, on average they are broader than Doppler boosting of an isotropic distribution in the spine rest frame to the LF (red dashed line). Such big differences in lepton momentum distributions between the pure e-ion SBL and the e+e-ion sandwich SBL are easy to understand. In both cases, leptons are accelerated by EM fields created by the difference in lepton and ion masses. In the pure e-ion case, the spine ions move with initial $\mathbf{p_{xL}}$=+451m$_i$c and carry all the initial energy. Their EM coupling to the spine electrons accelerates the electrons to the ion energy mainly via $\mathbf{E_x}$ fields induced in the spine, which leads to $\mathbf{p_{xL}} \gg \mathbf{p_y}$ and extreme beaming (Fig.10a). On the other hand, in the e+e-ion sandwich case, the sheath ions are initially at rest and there is no ion in the fast-moving spine. EM fields created at the SBL eventually accelerate some sheath ions via energy transfer from the spine leptons, which move with initial $\mathbf{p_{xL}}$=+451$m_e$c and carry all the initial energy. But the accelerated sheath ions can only reach low energy ($\mathbf{p_{xL}} \sim$ few $m_i$c) due to their large mass, and the low ion energy limits the energy reachable by the leptons. At the same time, the $\mathbf{E_x}$ fields induced in the spine are now much weaker, leading to broader lepton beaming (Fig.10b). Observationally, we therefore expect pure e-ion SBLs to emit much harder radiation with very narrow beaming and rapid time variability (due to variation of the observer view angle relative to the SBL orientation). In contrast, an e+e-spine interacting with an e-ion sheath should result in softer radiation, broader beaming and slower time variability. Such observational differences should be testable, especially for GRB jets where $\Gamma$ is large. Our PIC results also predict a stronger anti-correlation between lepton energy and beam angle for the pure e-ion SBL than for the e+e-ion sandwich SBL (Fig.10). We notice that while some electrons diffuse into the opposing flows (Fig.8b), positrons (Fig.9b) and protons do not diffuse into the opposing flows, due to the magnetic field barrier. While the main results presented here is based on runs with $\Gamma$ = 451 and therefore more relevant to GRBs, we have performed other PIC simulations with $\Gamma$ = 10's, more relevant to blazars. The results of those runs are qualitatively similar to those presented here. Maximum particle energies and beam patterns both scale with the Lorentz factor as expected.

In summary, our PIC simulation results show that efficient lepton acceleration up to $\gamma_e \sim \Gamma_i$ $m_i/m_e$ occurs in relativistic shear layers, and proceeds in a strongly anisotropic manner. In the case of a fast electron-positron spine surrounded by a slow or stationary electron-ion sheath, the spine-sheath interaction leads to moderate acceleration of the sheath ions to $\Gamma_i \sim$ few, and a broadening of the angular distribution of spine leptons compared to a direct Doppler boosting of an isotropic lepton distribution in the co-moving frame of the spine. However, the highest-energy leptons with Lorentz factors of $\gamma_e >$ a few $10^3$ retain the typical $\sim 1/\Gamma$ beaming characteristic in the laboratory frame. This has important consequences for the observable radiation. Keeping in mind that in the process of Compton scattering by relativistic leptons, the scattered, high-energy photon emerges in the direction of the scattering lepton. Hence, if shear-layer acceleration in relativistic astrophysical jets is the dominant acceleration mechanisms of relativistic leptons, jets viewed in the forward direction will exhibit very hard radiation spectra, much beyond the usual spectral hardening effect due to bulk Doppler boosting of a co-moving isotropic particle distribution (which is just a shift of the peak frequency by factor $\Gamma$). On the other hand, jets viewed at substantial off-axis angles (assuming that the shear layer is largely parallel to the global jet axis) will be dominated by the radiation from the broadly beamed low-energy particles, and thus exhibit rather soft spectra.

We emphasize that this beaming effect becomes much more pronounced in the case of a pure electron-ion shear layer. In the latter case, the highest energy electrons are beamed into a very narrow cone, several orders of magnitude narrower than the $\sim 1/\Gamma$ beaming characteristic of bulk Doppler boosting, while lower-energy particles still exhibit significant beaming. The narrow beaming may also explain the minute-scale rapid time-variability of some blazars (Tavecchio and Ghisellini, private communications).

The strong energy-dependence of the angular distribution of the relativistic leptons for spine-sheath jets has important consequences for the high-energy emission from relativistic jet sources, such as blazars and gamma-ray bursts. Essentially all current works on modeling the spectral energy distributions (SEDs) of blazars based on leptonic radiation scenarios assume a relativistic electron distribution which is isotropic in the co-moving frame of the bulk motion of the jet. In such models, very often quite high minimum Lorentz factors of the relativistic electron distribution have to be assumed (e.g., Ghisellini et al. 2010; Böttcher et al. 2013), which often appear to be rather arbitrarily chosen. Yet our PIC results naturally predict such low energy cutoffs (Fig.6d). Such a low-energy cutoff is necessary to reproduce the observed hard x-ray – gamma-ray spectra of blazars. This is particularly relevant for the spine-sheath model of Ghisellini et al. (2005) and Tavecchio & Ghisellini (2008). In this model, the radiative interaction between the spine and the sheath is proposed to increase the radiative efficiency due to relative Doppler boosting of the radiation fields produced in the two jet components. In this scenario, obviously, the shear-layer acceleration mechanisms investigated here becomes relevant. However, Ghisellini et al. (2005) and Tavecchio & Ghisellini (2008) assumed isotropic electron distributions in the respective co-moving frames, and also required large minimum electron Lorentz factors, at least in the spine. The differential beaming of leptons in the laboratory frame resulting from our PIC simulations provides a natural explanation for such hard spectra for blazars, in which we view the jet at a small angle from the jet axis. A detailed investigation of the resulting radiation characteristics based on PIC data (Hededal and Norlund 2005, Sironi and Spitkovsky 2009) is beyond the scope of the current paper and will be studied in follow-up works.

A problem similar to the low minimum electron Lorentz factors in blazars discussed above exists also for GRBs: The prompt hard x-ray / soft gamma-ray spectra of many GRBs show very hard low-energy slopes, even harder than a self-absorbed synchrotron spectrum (Rybicki and Lightman 1979), which have been considered as an argument against a synchrotron origin of the high-energy emission of GRBs (the "line of death" problem, Preece et al. 1999). This in turn stimulated a discussion on possible "photospheric" (optically thick) emission components in GRB spectra (e.g., Ryde 2005, Ryde et al. 2006). The extreme angle dependence of the lepton spectra found in our studies may also provide a solution for such unexpectedly hard x-ray – gamma-ray spectra.

The two cases of electron-positron versus electron-ion spine compositions differ markedly in the beaming characteristics of high-energy leptons. In the case of an electron-positron spine, even the highest-energy leptons will not exhibit a beaming pattern narrower than $\sim 1/\Gamma$, and substantial high-energy emission is still expected to be observable under moderate viewing angles characteristic of radio galaxies. In the case of GRB jets, this scenario predicts a substantial chance of detecting gamma-ray emission from GRBs at off-axis angles $\theta_{obs} > 1/\Gamma$. On the contrary, if the jet composition is pure electron-ion plasma in both the spine and the sheath, one expects a small population of sources viewed under very small viewing angles with $\theta_{obs} \ll 1/\Gamma$, with very hard gamma-ray spectra (such as those detected by LAT onboard Fermi Observatory), while a much larger population of off-axis sources with $\theta_{obs} > 1/\Gamma$ appear to have softer spectra. The recently emerging class of extreme BL Lac objects (e.g., Bonnoli et al. 2015) may possibly represent the small population of extremely narrowly beamed, very-hard-spectrum blazars expected in the electron-ion shear acceleration scenario.

**Acknowledgements:**

At Rice University this work was partially supported by NSF AST1313129 and Fermi Cycles 4 & 5 GI grants. The work of M.B. is supported through the South African Research Chairs Initiative (SARChI) of the Department of Science and Technology and the National Research Foundation[1] of South Africa under SARChI Chair grant no. 64789. We thank Prof. Tony Arber of the University of Warwick for providing the EPOCH code, and Drs. Fabrizio Tavecchio and Gabriele Ghisellini for useful discussions. Simulations with the EPOCH code were supported by the Data Analysis And Visualization Cyberinfrastructure at Rice University, funded by NSF under grant OCI-0959097. Simulations with the Zohar code were supported by the Lawrence Livermore National Laboratory.

REFERNCES

Alves, E.P., T Grismayer, SF Martins, F Fiúza, RA Fonseca, LO Silva, 2012, Ap. J. Lett. 746, L14.

Alves, E. P., Griesmayer, T., Fonseca, R. A., & Silva, L. O. 2014, New J. Phys., 16, 035007.

Alves, E. P., Griesmayer, T., Fonseca, R. A., & Silva, L. O. 2015, Phys. Rev. E, 92, 021101.

Birdsall, C. and A. B. Langdon, 1991 *Plasma Physics via Computer Simulation*, (IOP, Bristol).

Boettcher, M., 2007, Ap. Sp. Sci., 309, 95.

Boettcher, M., Reimer, A., Sweeney, K., & Prakash, A., 2013, ApJ, 768, 54

Bonnoli, G., Tavecchio, F., Ghisellini, G., & Sbarrato, T., 2015, MNRAS, 451, 611

Boyd, T. and Sanderson, J. 2003, *Physics of Plasmas* (Cambridge, UK).

Brady, C., Bennett, K., Schimtz, H., Ridgers, C. 2012, EPOCH Users Manual v4.0 July 29, 2012 (U. of Warwick, UK).

Chandrasekhar, S. 1981, *Hydrodynamic and Hydromagnetic Stability* (Dover, NY).

Ghisellini, G., Tavecchio, F., & Chiaberge, M., 2005, Ast. &Ap., 432, 401.

Ghisellini, G., Tavecchio, F., Foschini, L., Ghirlanda, G., Maraschi, L., & Celotti, A., 2010, MNRAS, 402, 497

Giroletti, M., G. Giovannini, L. Feretti, W. D. Cotton, P. G. Edwards, 2004, ApJ, 600, 127

Grismayer, T., E. P. Alves, R. A. Fonseca, and L. O. Silva, 2013, Phys. Rev. Lett. 111, 015005.

Gruzinov, A. 2008, arXiV:0803.1182.

Hededal, C. & Norlund, A., 2005, arXiV: astroph0511662.

Kargaltsev, O., Cerutti, B., Lyubarsky, Yu., & Striani, E., 2015, Space Sci. Rev. 191, 391.

Langdon, A.B. and B. Lasinski, 1976, Meth. Comp. Phys.16, 327, ed. B. Alder et al.

Lapenta, G. et al. 2007, Ap. J. 666, 949.

Liang, E., Boettcher, M., Smith, I., 2013a, Ap. J. Lett. 766. L19.

Liang, E., Fu, W., Boettcher, M., Smith, I., and Roustazadeh, P. 2013b, Ap. J. Lett. 779, L27.

Lyutikov, M. and Lister, M. 2010, ApJ 722, 197.

Maliani, Z., & Keppens, R., 2009, ApJ, 705, 1594

Meszaros, P. 2002, Ann. Rev. Ast. Ap. 40, 137.

Piran, T. 2004, Rev. Mod. Phys. 76, 1143.

Preece, R. D., Briggs, M. S., Mallozzi, R. S., Pendleton, G. N., Paciesas, W. S., & Band, D. L., 1998, ApJ, 506, L23

Rybicki, G. & Lightman, A. 1979, *Radiative Processes in Astrophysics* (Freeman, SF).

Ryde, F., 2005, ApJ, 625, L95

Ryde, F., Björnsson, C.-I., Kaneko, Y., Mészáros, P., Preese, R., & Battelino, M., 2006, ApJ, 652, 1400

Sironi, L., & Spitkovsky, A., 2009, ApJ 707, L92.

Tavecchio, F. & Ghsiellini, G., 2008, MNRAS, 385, L98

Weibel, S. 1959, Phys. Rev. Lett. 2, 83.

Yang, T. et al 1993, Phys. Fluids B 5, 3369.

Yang, T. et al 1994, Phys. Of Plasmas 1, 3059.

Yoon, P. 2007, Phys. Of  Plasmas 14, 024504.

Zhang, W., MacFadyen, A., and Wang, P. 2009, ApJ 692, L40.

Figure Captions

Fig.1 Setup of the (initially unmagnetized) shear flow PIC simulations of the e+e-ion sandwich. This paper focuses on the longitudinal P-mode evolution in the x-y plane only, since the transverse T-mode saturates at very low level compared to the P-mode. In the present case the plasma consists of right-moving e+e- pairs in the central 50% of the y-grid, referred to as the "spine", sandwiched between left-moving e-ion plasmas at the top 25% and bottom 25% of the y-grid, referred to as the "sheath". The simulation box has periodic boundary conditions on all sides. Inset: Sketch illustrating d.c. magnetic field creation by the ECCI. Throughout this paper and in all Figures, spatial scales are in units of the electron skin depth $c/\omega_e$.

Fig.2 (top row) Evolution of energy components for three homogeneous SBL runs with $p_o$=15: (a) pure e-ion plasma; (b) 90% e-ion + 10% e+e- plasma; (c) pure e+e- plasma. Here $E_{em}$ = EM field energy, $E_i$= ion energy, $E_e$=lepton energy. For cases (a)(b), only P-mode results are presented since T-mode has negligible effect. In case (c) we plot both P-mode and T-mode. In all cases $E_{em}$ saturates at $\sim$ 8% - 12% of total energy, much higher than fields amplified by KHI (Zhang et al 2009). In the two ion-dominated cases, the leptons reach energy equipartition with the ions after a finite time (from Liang et al 2013ab). (bottom row) $\mathbf{B_z}$ profiles of the same three runs at $t\omega_e$ = 3000: (d) pure e-ion plasma; (e) 90% e-ion + 10% e+e- plasma; (f) pure e+e- plasma. (blue and red denote opposite polarities, from Liang et al 2013ab).

Fig.3 Evolution of energy components of an e+e-ion sandwich SBL run: EM field energy (a, light-green dashed), positron energy (b, red), electron energy (c, green), ion energy (d, blue) and total energy (e, black). At late times the electron energy saturates at $\sim$ 25% of total energy, while the positron energy saturates at $\sim$ 4 % of total energy. The saturated EM field energy ($\sim$3% of total energy) is much lower than the homogeneous cases of Fig.2.

Fig.4 From top to bottom: spatial profiles of $\mathbf{B_z}$, $\mathbf{E_y}$, $\mathbf{j_x}$, $\mathbf{E_x}$, $\rho$ = net charge = $(n_+ - n_-)$, of the e+e-ion sandwich SBL run at three different times: $t\omega_e$ = (a) 1000, (b) 3000, (c) 10000. The $\mathbf{B_z}$ profiles at $t\omega_e$ = 3000 should be compared to the homogeneous SBL runs of Fig.2(d)(e)(f).

Fig.5 Density profiles n vs. y (averaged over x), for the four particle species of the e+e-ion sandwich SBL run at four sample times: $t\omega_e$ = (a) 1000, (b) 3000, (c) 10000, (d) 16000. Species :1 (red) denotes initially left-moving electrons in e-ion sheath, 2 (green) denotes initially right-moving positrons in e+e- spine, 3 (blue) denotes initially left-moving ions in e-ion sheath, 4 (magenta) denotes initially right-moving electrons in e+e- spine (3 and 4 are switched in (b)). These density profiles highlight the net charge distribution (black) as well as the diffusion of electrons into the opposing flows. At late times ($t\omega_e \geq 10000$) the e+e- spine expands against the e-ion sheath.

Fig.6 (a) Evolution of the total electron distribution function f($\gamma$) (particle no. per unit $\gamma$) vs. $\gamma$ for the e+e-ion sandwich SBL shows that the electron spectrum continues to harden over time: $t\omega_e$ = 1000 (red); 3000 (blue); 10000 (green); 16000 (magenta); (b) Same as (a) for spine positrons showing that the positrons gain little energy after $t\omega_e \sim 10000$; (c) Same as (a) for sheath ions showing the deceleration of the ions over time; (d) Comparison of electron (red), positron (blue) and ion (green) distributions at $t\omega_e$=16000. We see that the electrons have developed a hard power law between $\gamma_{min} \sim 400$ and $\gamma_{max} \sim 20000$ with sharp cut-offs at both ends, plus a narrow peak at $\gamma \sim 10000$.

Fig.7 Time-lapse profiles of the e+e-ion sandwich run on a larger (1024x12288) grid using the EPOCH code (color bar has been renormalized between frames). Panels (a)-(c) show spatial profiles of the various physical variables at three sample times: $t\omega_e$ = (a) 3000, (b) 10000, (c)

20000. (d) Evolution of the $\mathbf{B_z}$ field profile with time shows the expansion of the e+e- spine against the e-ion sheath at late times and the growth of nonlinear EM waves in the e+e- spine. There is little interaction between the EM waves created by the two SBLs until $t\omega_e \sim 15000$.

Fig.8 Phase plots of spine electrons for the pure e-ion run of Fig.2a at $t\omega_e=10000$, after Lorentz boosting $\mathbf{p_x}$ to the "laboratory frame" in which the sheath is initially at rest: (a) $\mathbf{p_{xL}}$ vs. $\mathbf{p_y}$, (b) $\mathbf{p_{xL}}$ vs. $y\omega_e/c$. By this time, many of the spine electrons have diffused into the sheath corresponding to $y\omega_e/c <500$ and $y\omega_e/c >1500$, and are decelerated and thermalized. These "cross-over" electrons form the bow-shaped low-energy population at the left of Fig.(a), while the arrow-shaped high energy population corresponds to electrons remaining in the spine ($500<y\omega_e/c<1500$). Note the narrow beaming of the highest energy electrons in Fig.(a), which are accelerated up to $\sim 0.5$ TeV. The broadening of $p_y$ around $p_x \sim 5\times10^5$ of Fig.(a) is caused by mid-plane collisions, which is likely an artifact of the small PIC box.

Fig.9 Phase plots of spine positrons for the e+e-ion sandwich SBL run at $t\omega_e=10000$, after Lorentz boosting $\mathbf{p_x}$ to the "laboratory frame" in which the sheath is initially at rest: (a) $\mathbf{p_{xL}}$ vs. $\mathbf{p_y}$, (b) $\mathbf{p_{xL}}$ vs. $y\omega_e/c$. No spine positrons can diffuse into the sheath region. Compared to Fig.8, the "beaming" of the highest energy positrons is much weaker, and they are only accelerated up to $\sim 10$ GeV. Spine electron phase plots are similar.

Fig.10 Comparison of the distribution of the tangent of the "beam angle" ($=|p_y/p_{xL}|$) vs. Lorentz factor $\gamma_L$ in the "laboratory frame", at $t\omega_e=10000$: (a) for the spine electrons of the pure e-ion SBL; (b) for the spine positrons of the e+e-ion sandwich SBL. We see that in case (a) most of the high-energy spine electrons (i.e. those that did not cross over to the sheath and get decelerated) have beam angles smaller than $1/\Gamma = 2.2\times10^{-3}$ (red dashed line). But in case (b), most of the spine positrons have beam angles larger than $1/\Gamma = 2.2\times10^{-3}$ (red dashed line). This means that the positron beam is broader than that from $\Gamma$ boosting of an isotropic lepton distribution in the spine rest frame.

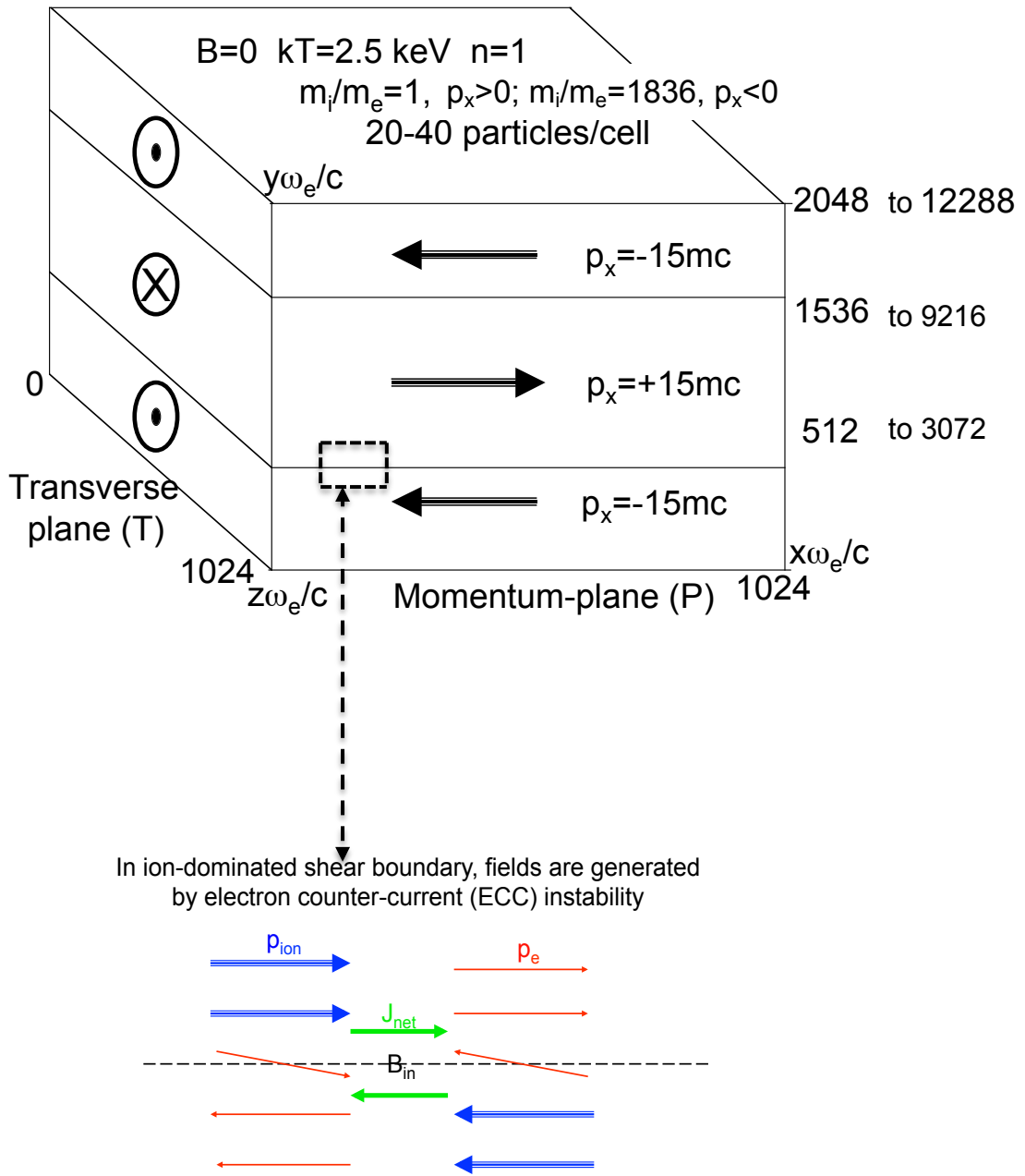

In ion-dominated shear boundary, fields are generated by electron counter-current (ECC) instability

Fig.1

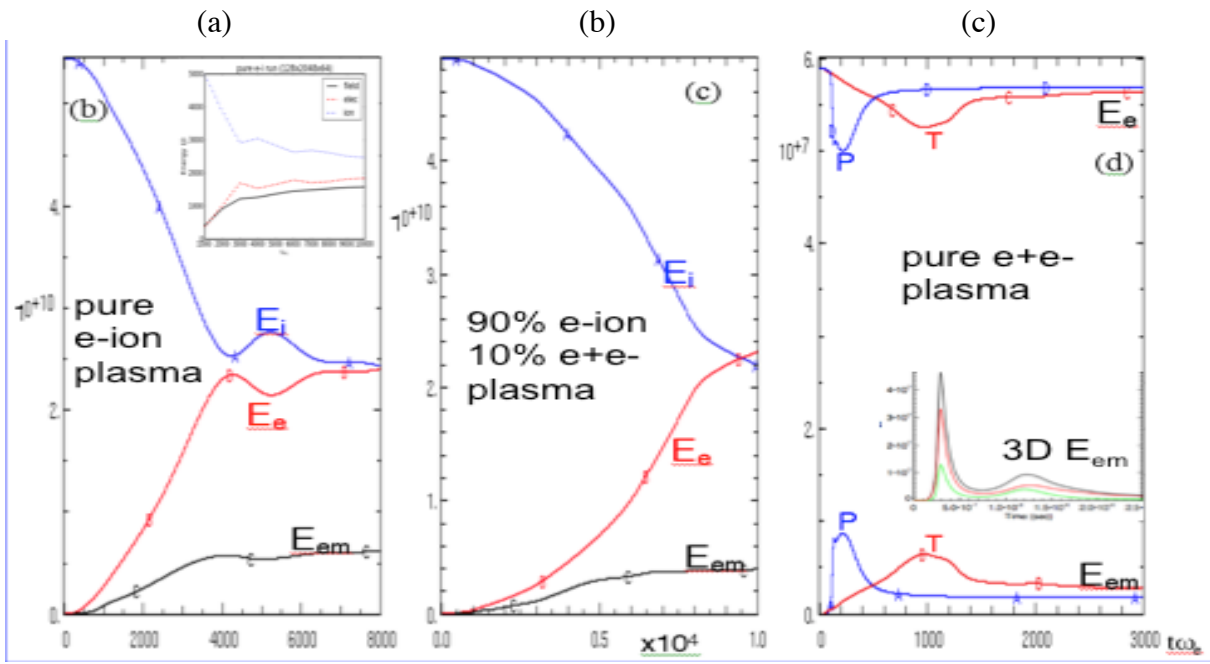

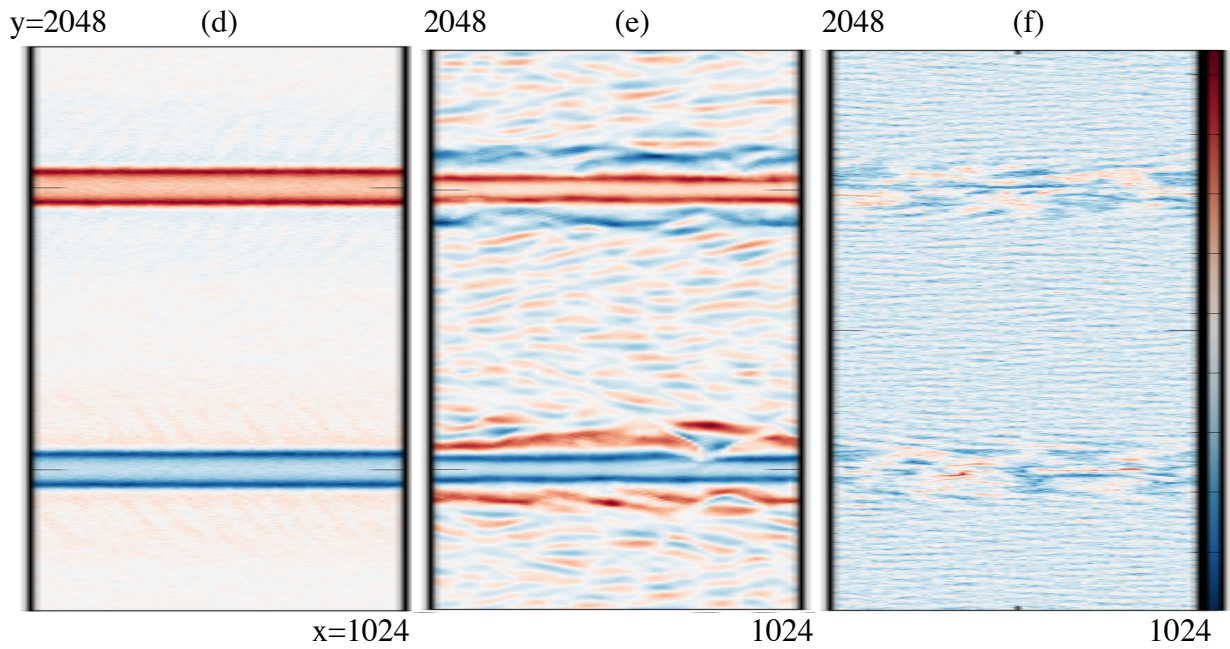

Fig.2

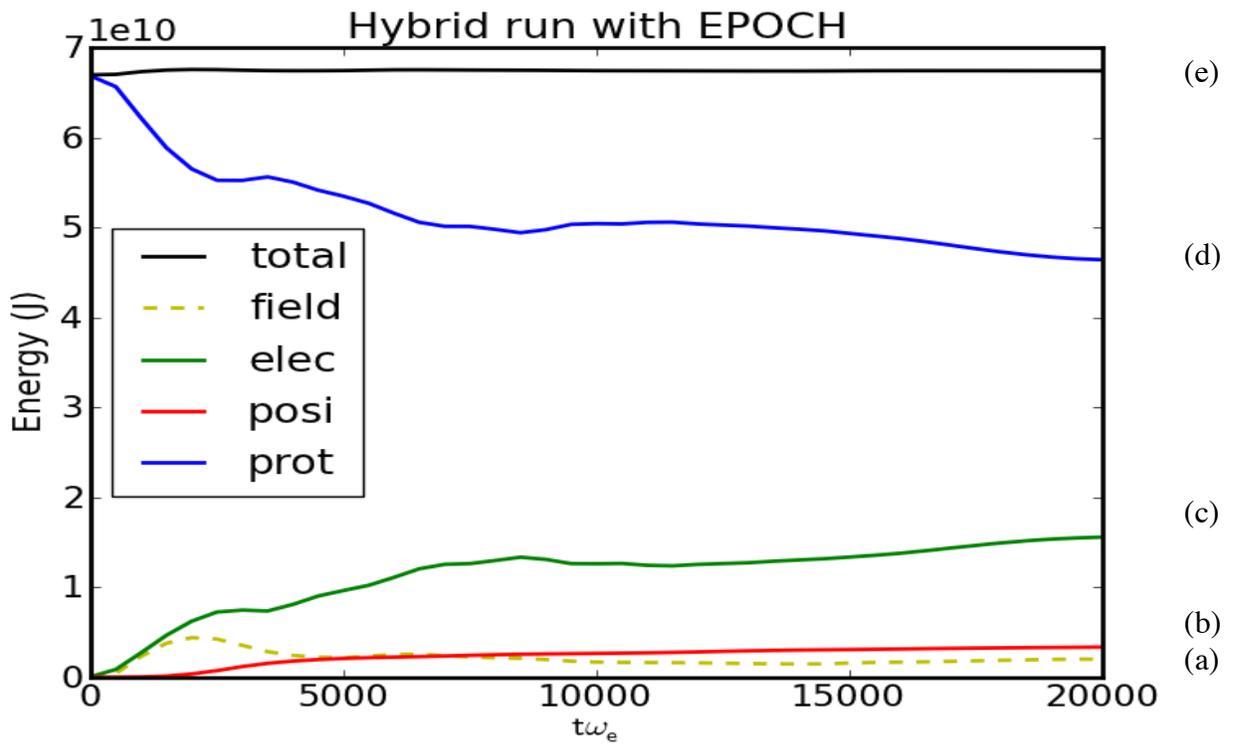

Fig. 3.

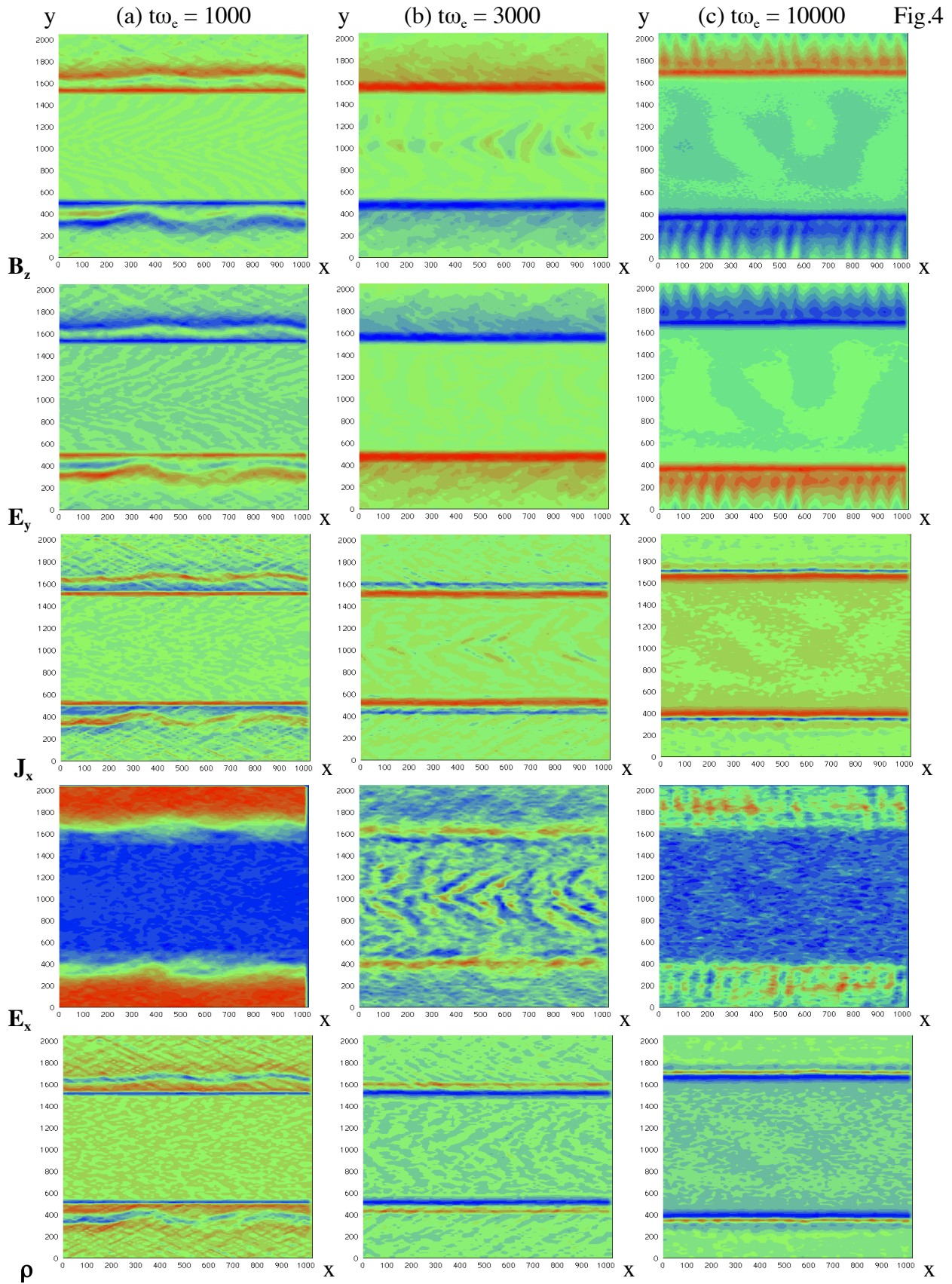

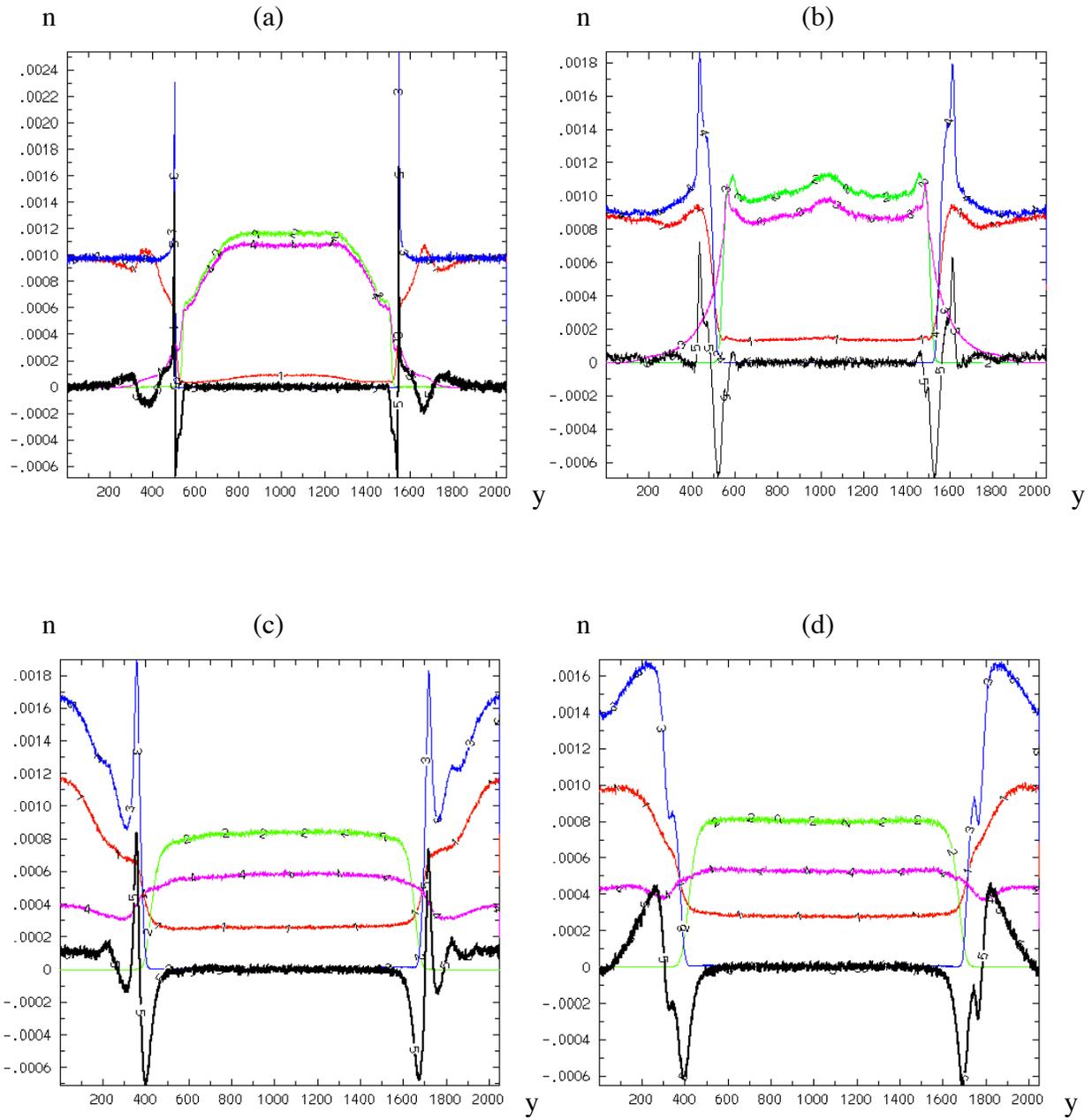

Fig.5

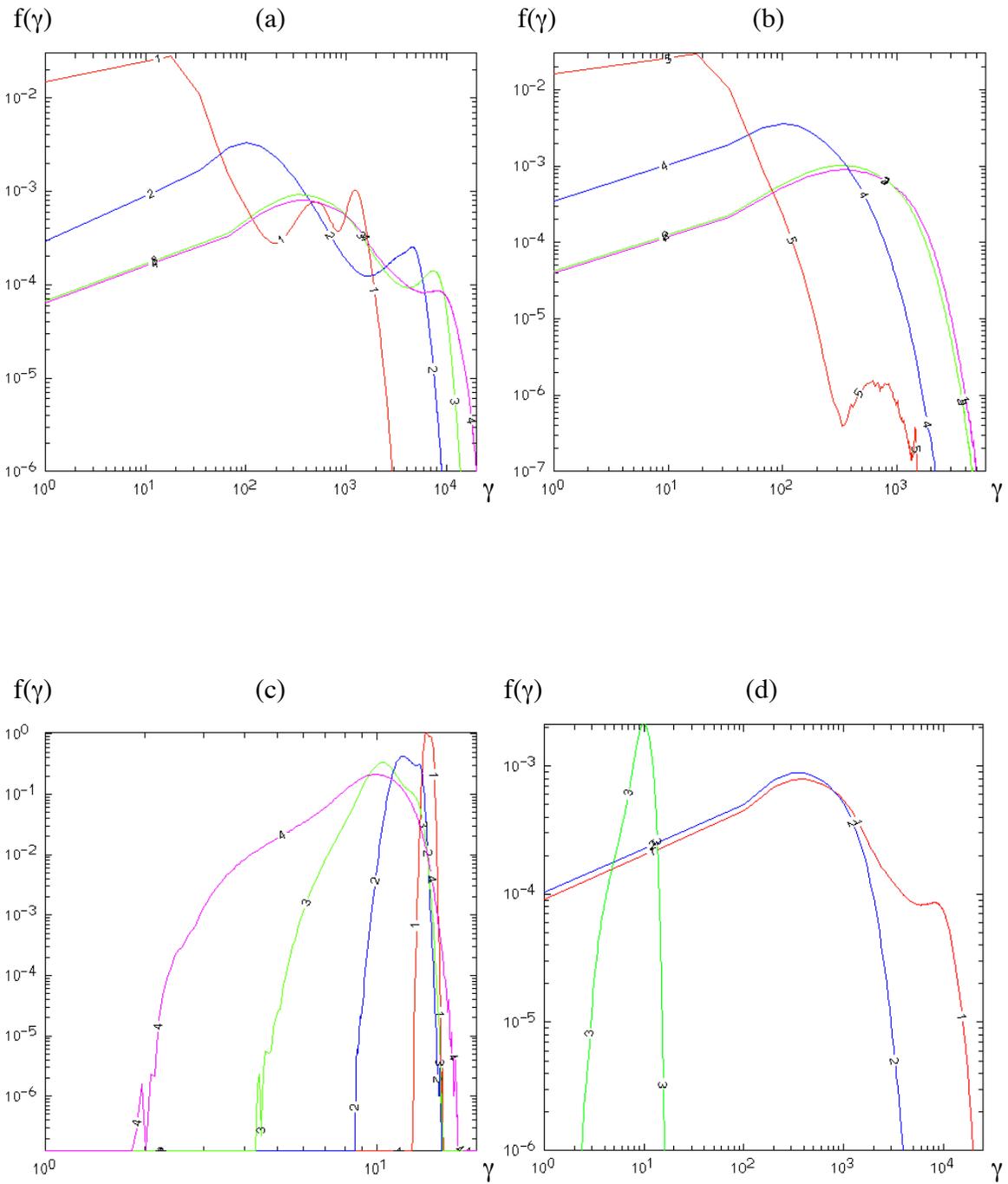

Fig.6

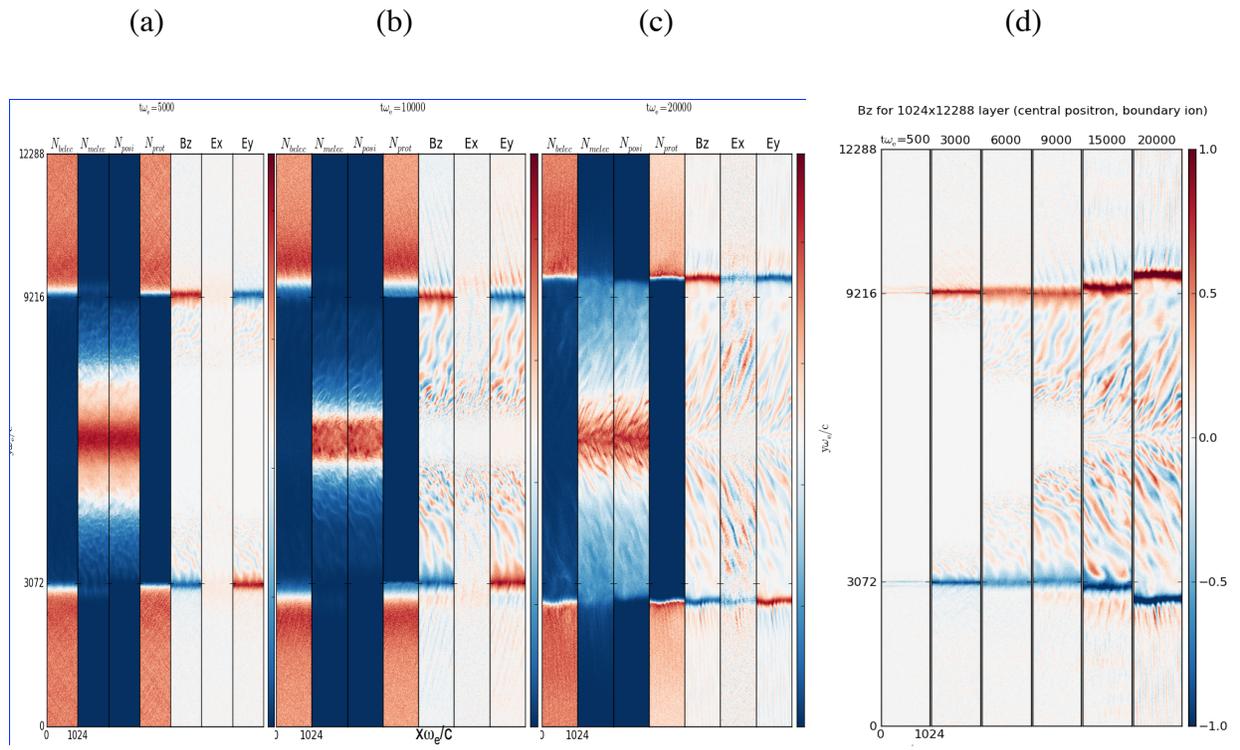

Fig.7

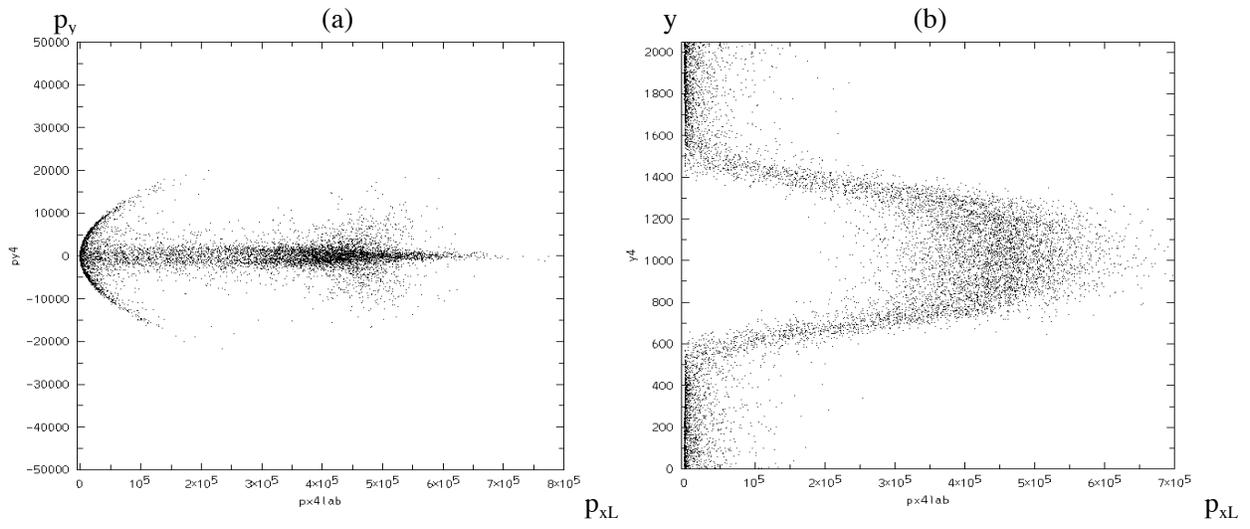

Fig.8

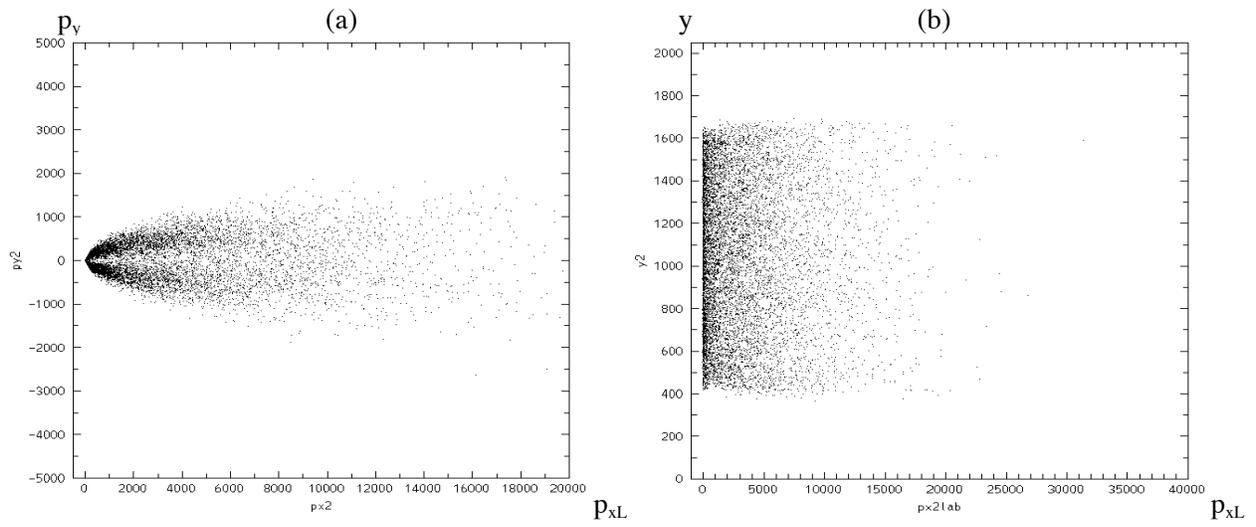

Fig.9

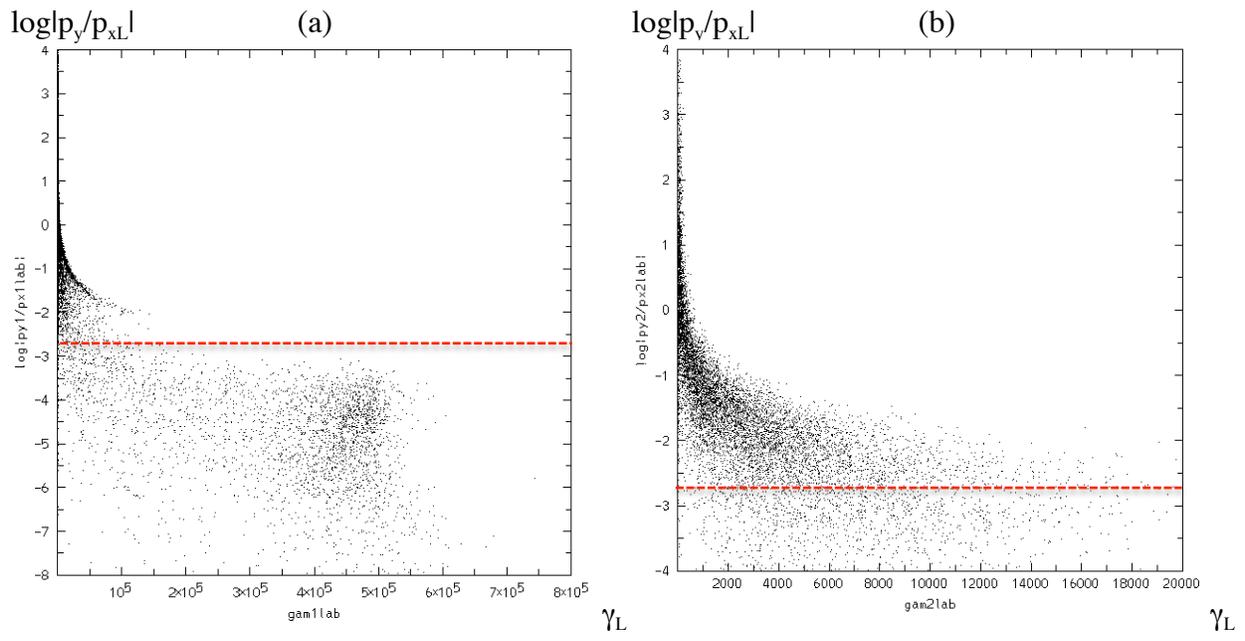

Fig.10